\documentclass[journal=jacsat,manuscript=article]{achemso}

\usepackage{chemformula} 
\usepackage[T1]{fontenc} 
\usepackage{graphicx}
\usepackage{caption}
\usepackage{subcaption}

\graphicspath{{./figures/}}

\author{Kiana Malmir}
\affiliation[Oxford University]
{Department of Materials, University of Oxford, Parks Road, Oxford OX1 3PH, United Kingdom}

\author{William Okell}
\affiliation[Oxford University]
{Department of Materials, University of Oxford, Parks Road, Oxford OX1 3PH, United Kingdom}

\author{Aurélien A P  Trichet}
\affiliation[Oxford HighQ]
{Oxford HighQ Ltd, Centre for Innovation and Enterprise, Begbroke Science Park, Oxford OX5 1PF, United Kingdom}
\author{Jason M Smith}
\email{jason.smith@materials.ox.ac.uk}
\affiliation[Oxford University]
{Department of Materials, University of Oxford, Parks Road, Oxford OX1 3PH, United Kingdom}
\alsoaffiliation[Oxford HighQ]
{Oxford HighQ Ltd, Centre for Innovation and Enterprise, Begbroke Science Park, Oxford OX5 1PF, United Kingdom}

\title[An \textsf{achemso} demo]
  {Characterisation of Nanoparticle Size Distributions in a Fluid using Optical Forces
  }

\abbreviations{IR,NMR,UV}

\begin{document}








\begin{abstract}
  We introduce a method for analyzing the physical properties of nanoparticles in fluids via the competition between viscous drag and optical forces. By flowing particles through a microfluidic device containing an optical microcavity which acts as a combined optical trap and sensor, the variation of the rate of trapping events with the different forces can be established. A clear threshold behaviour is observed which provides a measure of a parameter combining the dielectric polarizability and the hydrodynamic radius. This technique could be applied in combination with other analytic techniques to provide a detailed physical characterisation of particles in solution.
\end{abstract}

 \begin{figure*}
\centering
    \includegraphics[width=5.5in]{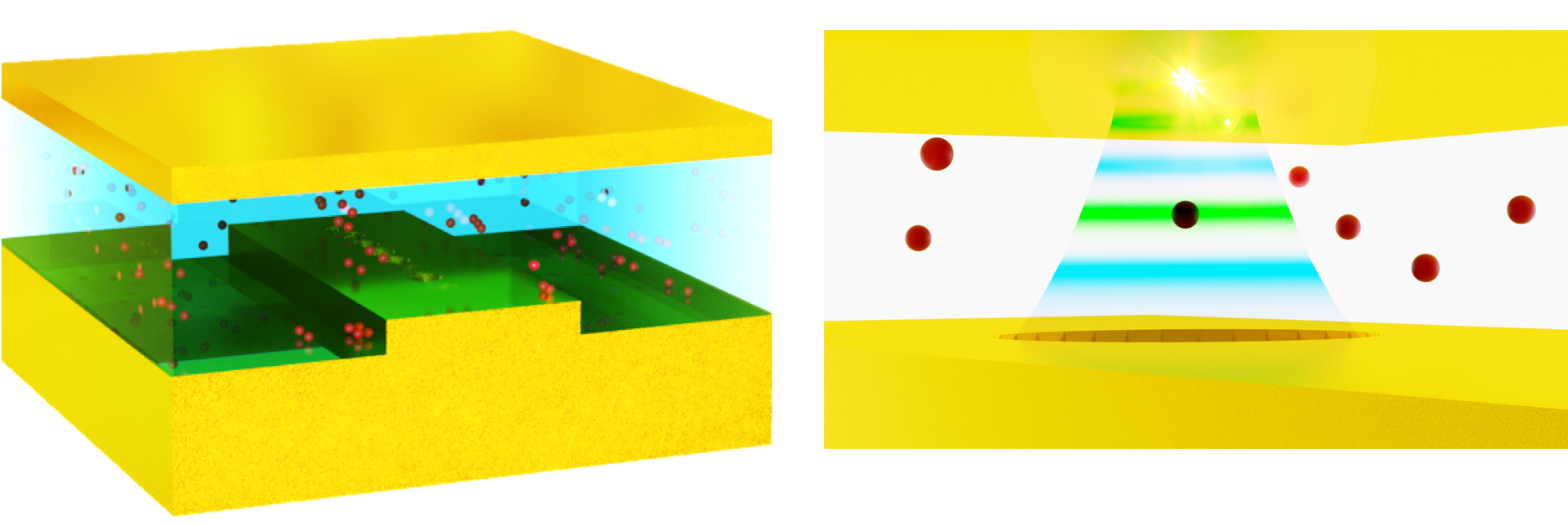}
\caption{Schematic of the characterisation instrument which incorporates optical microcavities within a microfluidic flow cell such that individual nanoparticles passing through the cavity mode can be detected and subjected to optical forces.}
\label{fig:featured}
\end{figure*}

\section{Introduction}
The characterization of nanoparticles in solution is of increasing importance for applications in biomedical, environmental, and materials sciences \cite{barrios2012integrated,chen2012sensitive,malmir2016ultrasensitive,malmir2020ultrasensitive}. In particular, the engineering of nanoparticles for use in medical applications such as drug delivery requires accurate characterization of the physical and chemical characteristics of particles within the fluid medium. For instance, the measurement of the loading of lipid nanoparticles with messenger RNA (mRNA) is of increasing importance for gene therapy and vaccine development \cite{swingle2021lipid}.

Particle size and size distribution are readily measured using light scattering techniques. Dynamic Light Scattering (DLS)\cite{stetefeld2016dynamic} is fast and sensitive to particles as small as 5 nm, but provides limited resolution for polydisperse samples where larger particles dominate the scattered signal \cite{soares2018nanomedicine,mourdikoudis2018characterization}. Nanoparticle Tracking Analysis (NTA) and tunable resistive pulse sensing (TRPS) provide single particle measurements of particle size allowing for more reliable measurement of size distributions. \cite{surrey2012quantitative,orlova2011structural,seo2007atomic,nakamura2017transmission,filipe2010critical,troiber2013comparison,bell2013quantitation}. 

Beyond particle size, centrifugation techniques allow direct measurement of the mass density of particle ensembles \cite{shard2018measuring,minelli2018measuring}, and new optical techniques have recently emerged which measure the composition of individual particles via the dielectric polarizability, a parameter which depends on both particle volume and dielectric constant and can in some cases provide a proxy for particle mass. Such measurements are made more challenging by the rapid Brownian motion of the particles in the bulk fluid, and various approaches have been taken to ensure that reliable quantitative measurements can be made. Techniques that allow measurement of other parameters, and in particular those that allow measurement of multiple parameters on individual particles so that correlations can be observed, remain of significant interest.   

In this work, we introduce a new method for characterizing the nanoparticles in a fluid by measuring the competition between optical gradient forces due to a focused laser mode and viscous drag of a flowing fluid. We find that varying the relative strengths of these forces and counting the number of particle trapping events provides a means for measuring the distribution of a parameter related to the polarisability which includes size and composition information. This parameter can be used to extract the particle size distribution for nanoparticles of known composition, or to extract both size and composition information in combination with a complementary characterisation technique.

The equation of motion for a spherical particle with polarisability $\alpha$ in a viscous fluid flowing at uniform velocity $\underline{v}_0$ and illuminated with an optical field of intensity distribution $I\left(\underline{r}\right)$ is
\begin{align}\label{Brownian}
 \begin{aligned} 
\gamma \left( \frac{d\underline{r}}{dt} - \underline{v}_0 \right) =\frac{1}{2n_m \epsilon_0 c}\alpha \nabla I\left(\underline{r}\right) +\sqrt{2K_{B} T \gamma}~\underline{W}\left(t\right). 
\end{aligned}
\end{align}
Here $\gamma=6\pi\eta a$ is the coefficient of friction where $\eta$ is the viscosity and $a$ is the particle radius, so that the left hand side of the equation represents the viscous drag force acting on the particle. The first term on the right hand side is the optical force in the dipole approximation where $n_m$ is the refractive index of the surrounding medium, and the second term represents the Brownian force acting on the particle in which $K_{B}T$ is the thermal energy and $\underline{W}\left(t\right)$ is a time-varying normally distributed random vector \cite{volpe2013simulation}.
Within the dipole approximation the polarisability is 
\begin{equation}\label{E:alpha}
    \alpha=4\pi a^3 n_m^2 \epsilon_0 \frac{m^2-1}{m^2+2},
\end{equation}
where $m$ is the ratio of refractive index of the particle ($n_p$) to that of the surrounding medium ($n_m$), i.e. $m=n_p/n_m$. It can be seen from equations \eqref{Brownian} and \eqref{E:alpha} that for a given optical intensity distribution $I\left(\underline{r}\right)$ and flow velocity $\underline{v}_0$, the balancing of the optical force with the drag force due to the flowing fluid is achieved for a threshold value, $\beta_T$, of the parameter 
\begin{equation}\label{beta}
    \beta=a^2\frac{m^2-1}{m^2+2},
\end{equation}
such that particles with $\beta>\beta_T$ may become trapped in the optical mode.

Apparatus which allows the rate of trapping events $\left(\Gamma_{trap}\right)$ to be detected as a function of the flow speed or optical power can therefore be used to determine the distribution $N\left(\beta\right)$ by differentiating the measured trapping rate with respect to $\beta_T$,
\begin{equation}\label{Nbeta}
    N\left(\beta\right)=-\frac{d\Gamma_{trap}}{d\beta_T}.
\end{equation}

 The thermal motion of the particle introduces a random element to the time a particle spends in the mode for values of $\beta$ close to the threshold and is therefore a source of broadening of the distribution function. 

\begin{figure}[!t]
\centering
\begin{subfigure}[b]{.46\textwidth}
    \centering
    \includegraphics[width=1\linewidth]{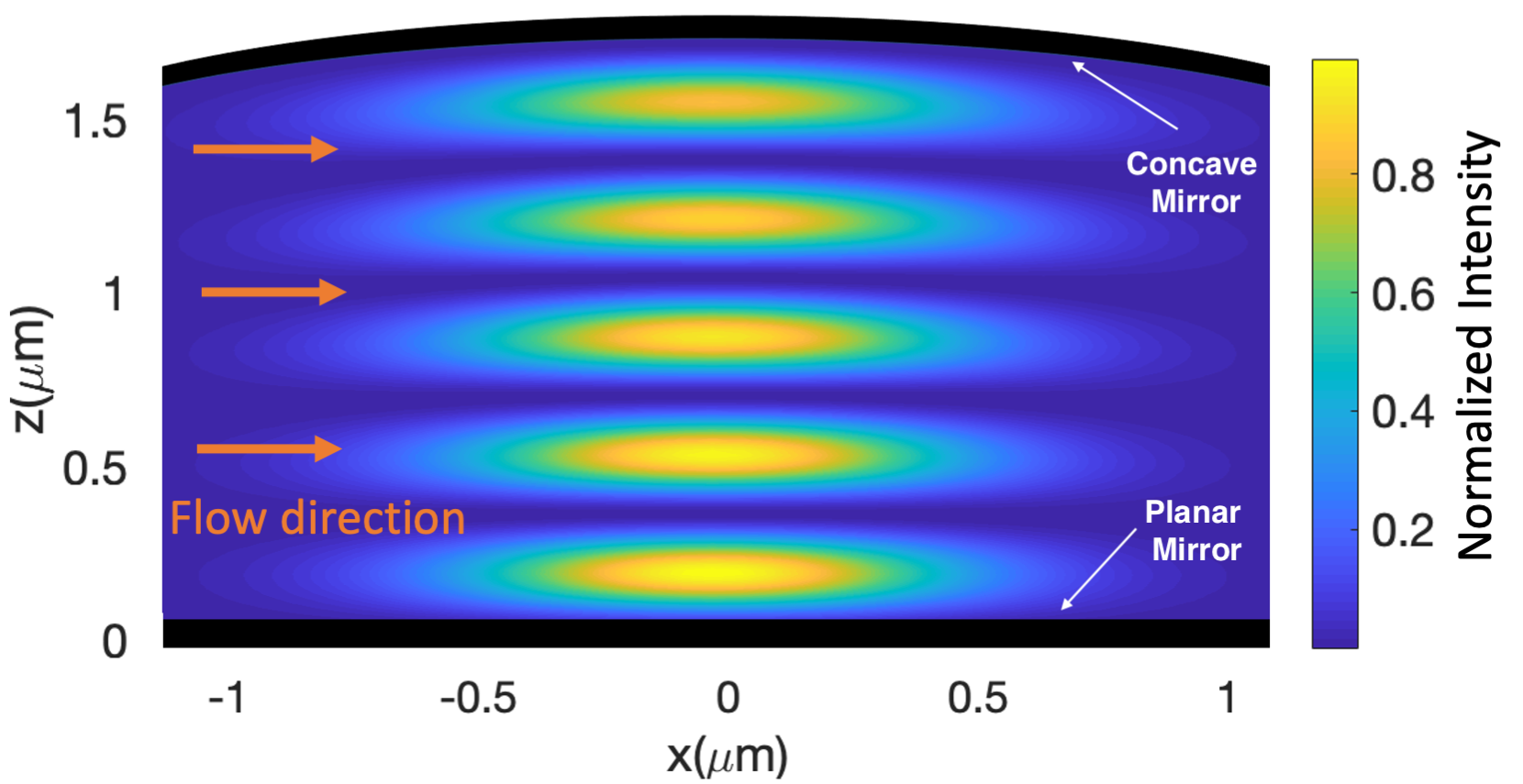}
    \caption{}	\label{fig:antinodes_flow}
\end{subfigure}
\begin{subfigure}[b]{.4\textwidth}
  \centering
  \includegraphics[width=1\linewidth]{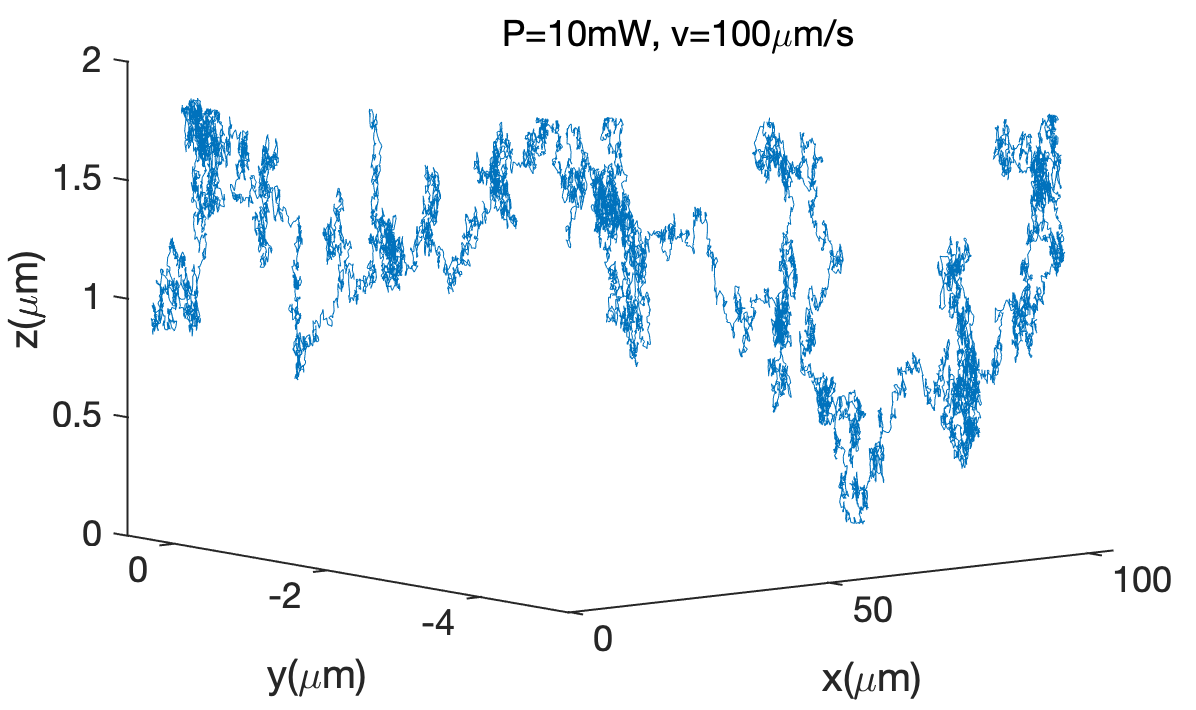}  
  \caption{}
  \label{fig:Traj2}
\end{subfigure}
\begin{subfigure}[b]{.4\textwidth}
  \centering
  \includegraphics[width=1\linewidth]{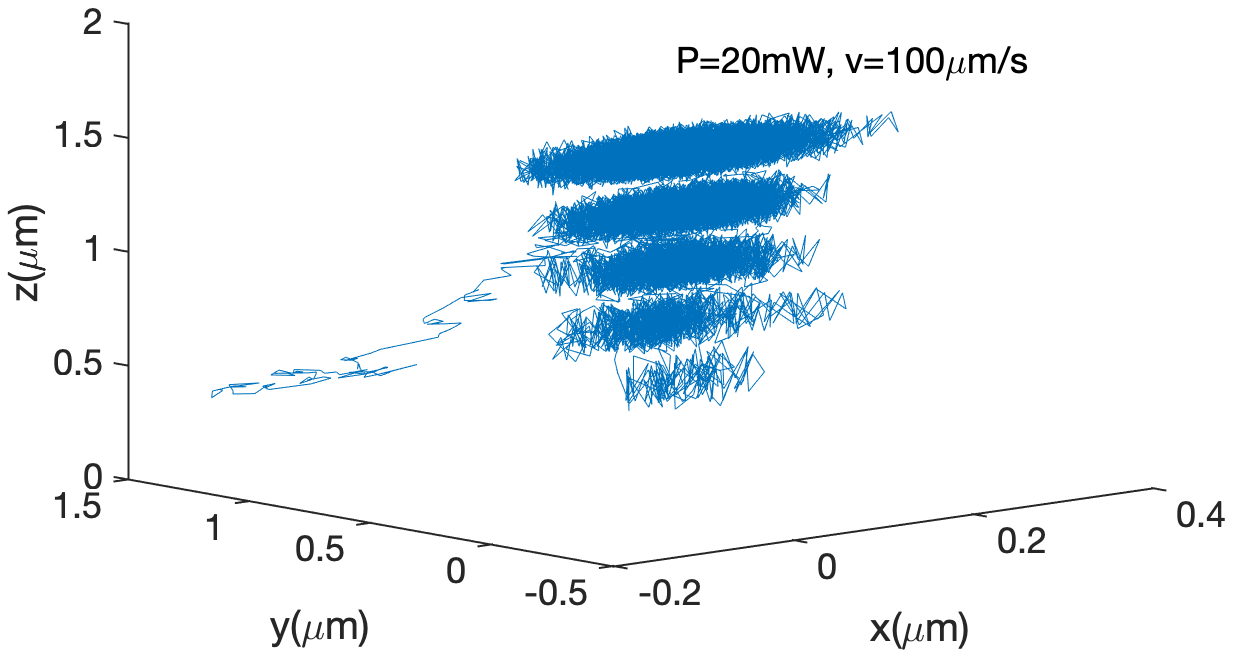}  
  \caption{}
  \label{fig:Traj3}
\end{subfigure}
\begin{subfigure}[b]{.4\textwidth}
  \centering
  \includegraphics[width=1\linewidth]{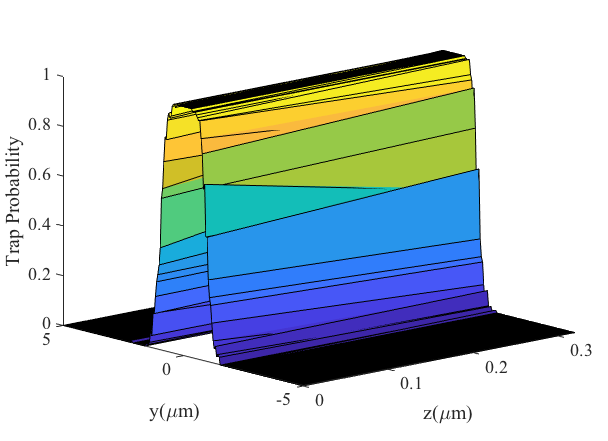} 
  \caption{}
  \label{fig:probab}
\end{subfigure}
\caption{(a) False colour-scale plot of the normalized optical intensity distribution $I\left(\underline{x,z}\right)$ in an open access microcavity with 5 optical anti-nodes between a planar and concave mirror. The intensity distribution is cylindrically symmetric about the $z$ axis, and the fluid flow direction is parallel to $x$. (b,c) Example simulated trajectories of a particle of radius 100~nm subject to the flow and optical forces described in the text. The flow speed is 100 $\mu$ m s$^{-1}$ and the intracavity power is 10 mW and 20 mW resepctively. (d) A typical trapping probability distribution as a function of the initial position of the particle in the $(y,z)$ plane.}
\label{fig:Traj_Prob}
\end{figure}

In this work, we generate the optical mode in a plano-concave open microcavity built into a microfluidic flow cell as described in \cite{trichet2016nanoparticle} and depicted in figure~\ref{fig:featured} such that $I\left(\underline{r}\right)$ can be expressed as a standing wave of a Gaussian beam with cylindrical symmetry about the axis

 \begin{equation}\label{E:I_R}
\begin{aligned}
I\left(\rho,z\right)=\frac{2P}{\pi\omega(z)^2}\sin^2\left(kz+\frac{k\rho^2}{2R\left(z\right)}-\arctan{\left(\frac{z}{z_R}\right)} \right)e^{\frac{-2\rho^2}{\omega(z)^2}}.
\end{aligned}
\end{equation}
Here, $\rho$ and $z$ are the radial and axial coordinates with their origin at the beam focus which corresponds to a node of the electric field on the planar mirror. $P$ and $k$  are the intracavity power and optical wavenumber, respectively. The parameters $z_R$, $w\left(z\right)$ and $R\left(z\right)$ are the parameters of the Gaussian beam that are established from the curvature of the concave mirror and the cavity length (see Supporting Information). A plot of the normalised intensity distribution for a resonant mode in a water-filled cavity at a wavelength of 640 nm, with $z_R$= 3.26 $\mu$m and with five anti-nodes between the two mirrors is shown in figure~\ref{fig:antinodes_flow}.
The use of a microcavity allows for detection of nanoparticles via shifts in the resonance, such that individual particles passing through the mode register as discrete events the duration of which can be recorded.

To simulate the measurement method we use a Monte Carlo model based on equation \eqref{Brownian} (see Supporting Information for detail). We define transverse axes $x$ and $y$ such that $\rho^2=x^2+y^2$ and select $x$ as the direction of fluid flow. Particles are `launched' starting from a position 1 $\mu$m upstream of the cavity mode ($x$ = -1~$\mu$m) with selected values of position ($y,z$), fluid flow rate $v$ and optical power $P$. Figure \ref{fig:Traj_Prob}(b,c) show example trajectories for a particle of radius 100~nm and with flow speed $v$ = 100~$\mu$m s$^{-1}$. Figure \ref{fig:Traj_Prob}b shows an example trajectory of a particle that is not trapped by the mode at $P$ = 10~mW, while figure \ref{fig:Traj_Prob}c shows a similar particle being trapped when $P$ is increased to 20~mW. To determine a trapping probability, trapping was defined to have occurred if the particle did not pass $x$ = 1~$\mu$m within a time equal to 10~$\mu$m divided by the flow speed.
For a given choice of $v$ and $P$ the initial position ($y,z$) was varied and the trapping probability (based on 100 repetitions per position) was established to produce a distribution map (figure \ref{fig:Traj_Prob}d). It was found that within the range of parameters used the trapping probability was independent of the initial $z$ position and so the trapping cross section $\sigma$ was defined by integrating the distribution map with respect to $y$ only. The trapping rate $\Gamma$ is then given by the product of the trapping cross section, the cavity length ($L$), the flow speed ($v$) and nanoparticle concentration per unit volume in the fluid ($C_{NP}$):
 \begin{equation}\label{E:Gamma}
\begin{aligned}
\Gamma=\sigma L v C_{NP}.
\end{aligned}
\end{equation}

\begin{figure}[!t]
\centering
  \includegraphics[width=1\textwidth]{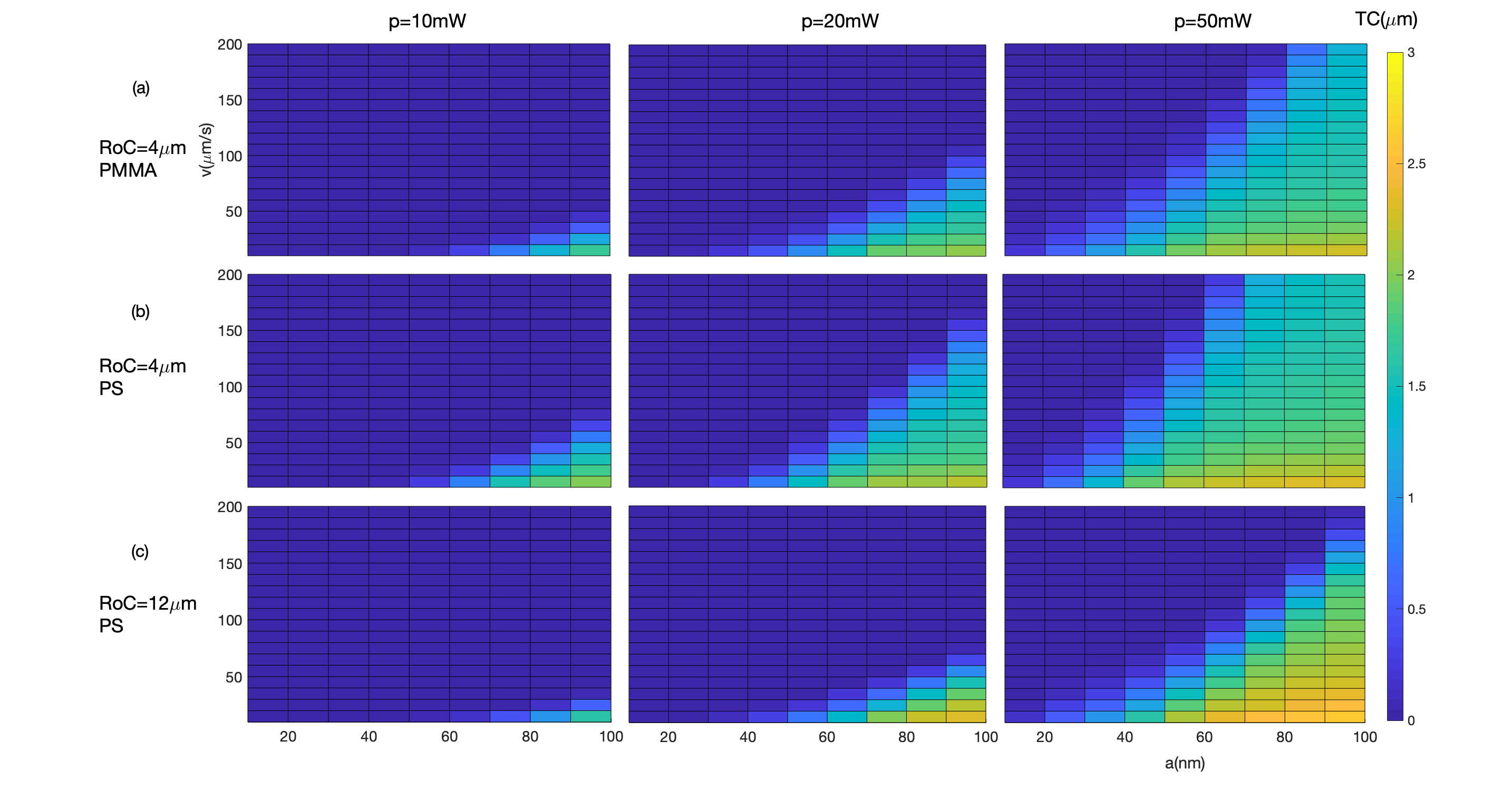}
  \caption{Trapping cross-section $\sigma$ as a function of type, radius and velocity of the particle in different situations such as optical beam waist, refractive index of particle. }
    \label{fig:trapcross}
\end{figure}

Figure~\ref{fig:trapcross} shows the simulated variation of $\sigma$ with several experimental parameters, as a series of false colour-scale plots. Each plot shows the dependence on flow speed $v$ and particle radius $a$; maps are generated for optical powers of $P$=10~mW, 20~mW and 50~mW (columns) and for two different mirror radii of curvature ($R_L$) and particle refractive index ($n_p$) (rows). The general result is as expected - that small particles in a rapidly flowing fluid (upper left of maps) do not trap, while larger particles in a slow flow speed show substantial trapping cross sections of order $2~\mu$m. The striking feature of these maps is that in each case the boundary between the region with no trapping and the region where trapping occurs is reasonably sharp, such that at a given flow speed $\sigma$ rises from zero to around $1.5~\mu$m within an increase in particle radius of about 10~nm. It is this sharp boundary that provide a basis for using the method for quantitative measurement.

\begin{figure}[!t]
\centering
\begin{subfigure}{.48\textwidth}
  \centering
  \includegraphics[width=1\linewidth]{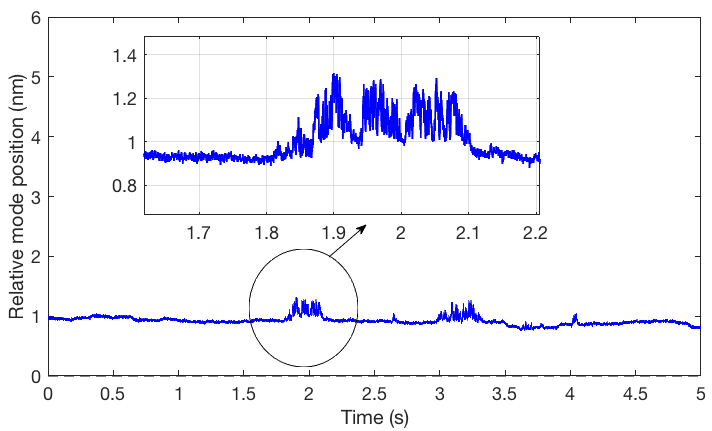} 
  \caption{}
  \label{fig:zoomed}
\end{subfigure}
\begin{subfigure}{.48\textwidth}
  \centering
  \includegraphics[width=1\linewidth]{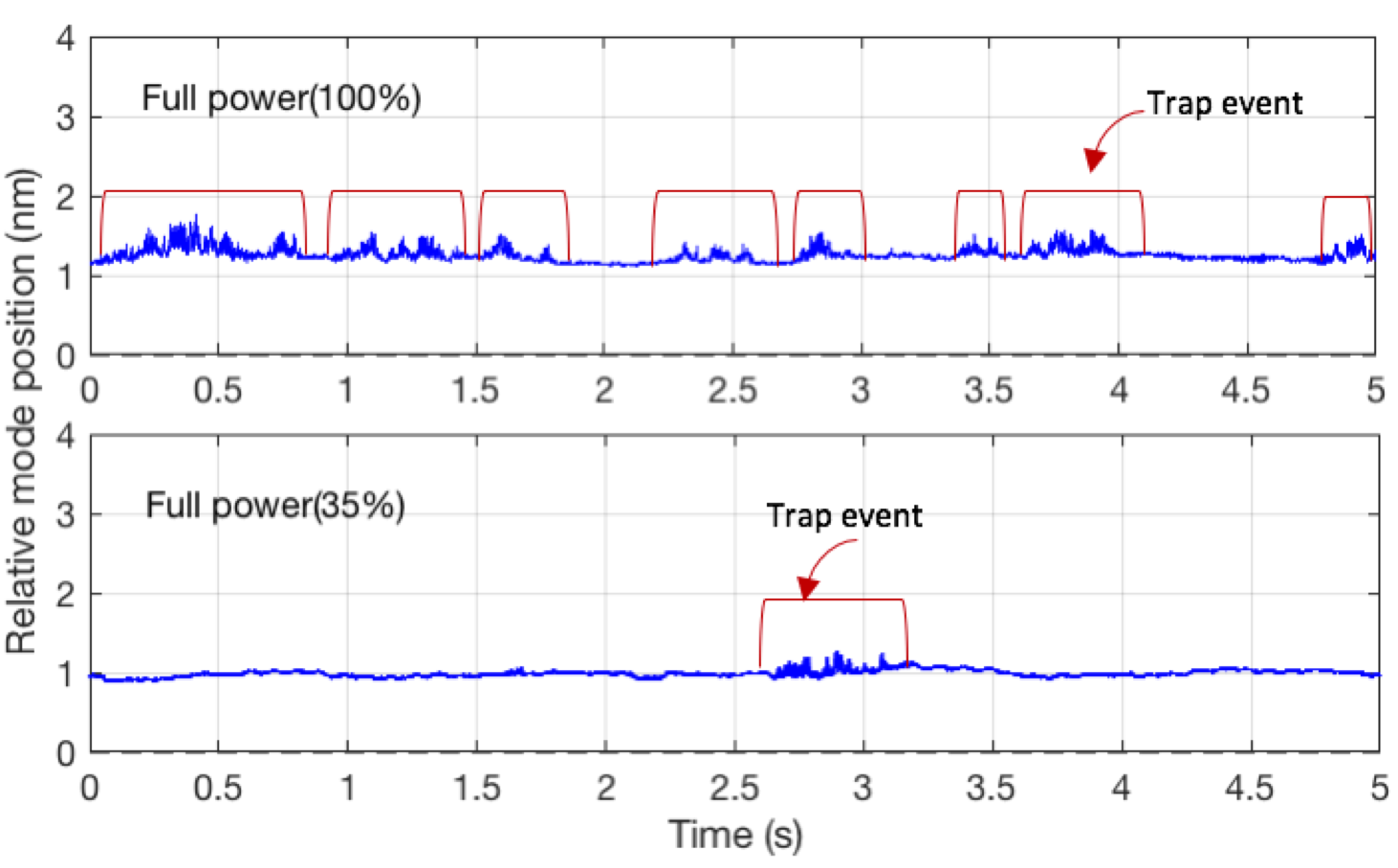}  \caption{}
  \label{fig:highlow_power}
\end{subfigure}
\caption{ Example experimental data showing single particle events. (a) Prolonged mode shift of a trapped particle (b) the effects of laser power on the number of trapping events.}
\label{fig:modeshift}
\end{figure}

The location of this boundary within the map depends on the values of the fixed parameters $P$, $w_0$ and $n_p$. As expected the boundary moves to smaller particle sizes and higher flow speeds for increasing $P$ and $n_p$ and for decreasing $w_0$. Within each plot the quadratic dependence of the threshold speed on particle radius given in equation \eqref{beta} can be seen, and the sensitivity to refractive index is shown by the comparison of PS and PMMA in figures \ref{fig:trapcross}a and \ref{fig:trapcross}b. The dependence on fixed experimental parameters indicates the ability to tune the measurement to different particles. The value of $\beta$ that balances the maximum trapping force with the flow force is found to be
 \begin{equation}\label{E:beta_thresh}
\begin{aligned}
\beta_T = \frac{3 \pi \eta v c w^3 \sqrt{e}}{2 n_m P}
\end{aligned},
\end{equation}
which reveals the scaling of the technique sensitivity with these experimental parameters.
In the experiments, nanoparticle solutions of concentration $10^{11}$~ml$^{-1}$ were pumped through a microfluidic flow cell with integrated microcavity measurement system. A constant differential pressure of 2~mbar was established using a hydrostatic flow regulator \cite{elve}, resulting in a peak flow speed of $\sim 50~\mu$m s$^{-1}$. Single nanoparticles passing through the microcavity register as discrete `events' the  duration of which is our primary indicator for trapping. Based on the flow speed alone we expect events due to particles that flow freely through the cavity mode to display durations of about 20~ms, and so we define a trapping event as one with duration exceeding 200~ms preceded by a period exceeding 100~ms with no observed mode shift. Figure \ref{fig:modeshift}a shows exemplar mode shift events highlighting both trapped and non-trapped particles. 

An important consideration for the pressure-driven flow used here is that the flow speed is not uniform but follows a parabolic profile across the flow channel cross section, with zero flow rate at the mirror surfaces. We select the mirror separation such that five anti-nodes of the optical field lie within the flow channel, whereby we expect about two-thirds of the recorded events to result from the anti-nodes nearest the centre of the flow channel. The parabolic flow profile is included in the simulation to ensure accurate comparison with experimental data.

The distribution $N(\beta)$ for particles in a solution is measured by sweeping the magnitude of the optical force via the laser power while maintaining a constant flow speed. Figure \ref{fig:modeshift}b shows broadly how the number of trap events can be seen to increase with increased laser power. 

\begin{figure}[!t]
\centering
\begin{subfigure}{.3\textwidth}
  \centering
  \includegraphics[width=1\linewidth]{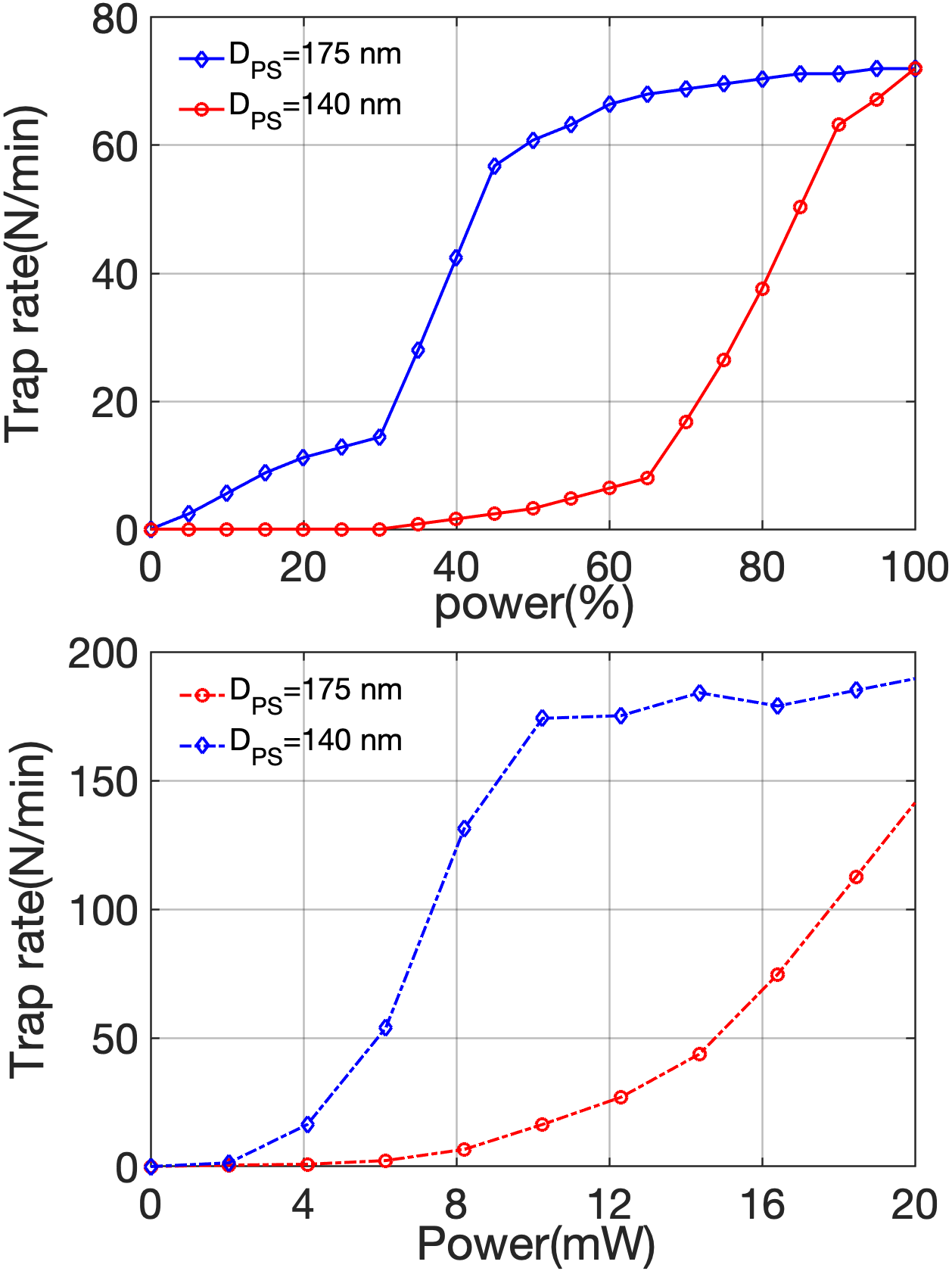}  
  \caption{}
  \label{fig:PS_T}
\end{subfigure}
\begin{subfigure}{.29\textwidth}
  \centering
  \includegraphics[width=1\linewidth]{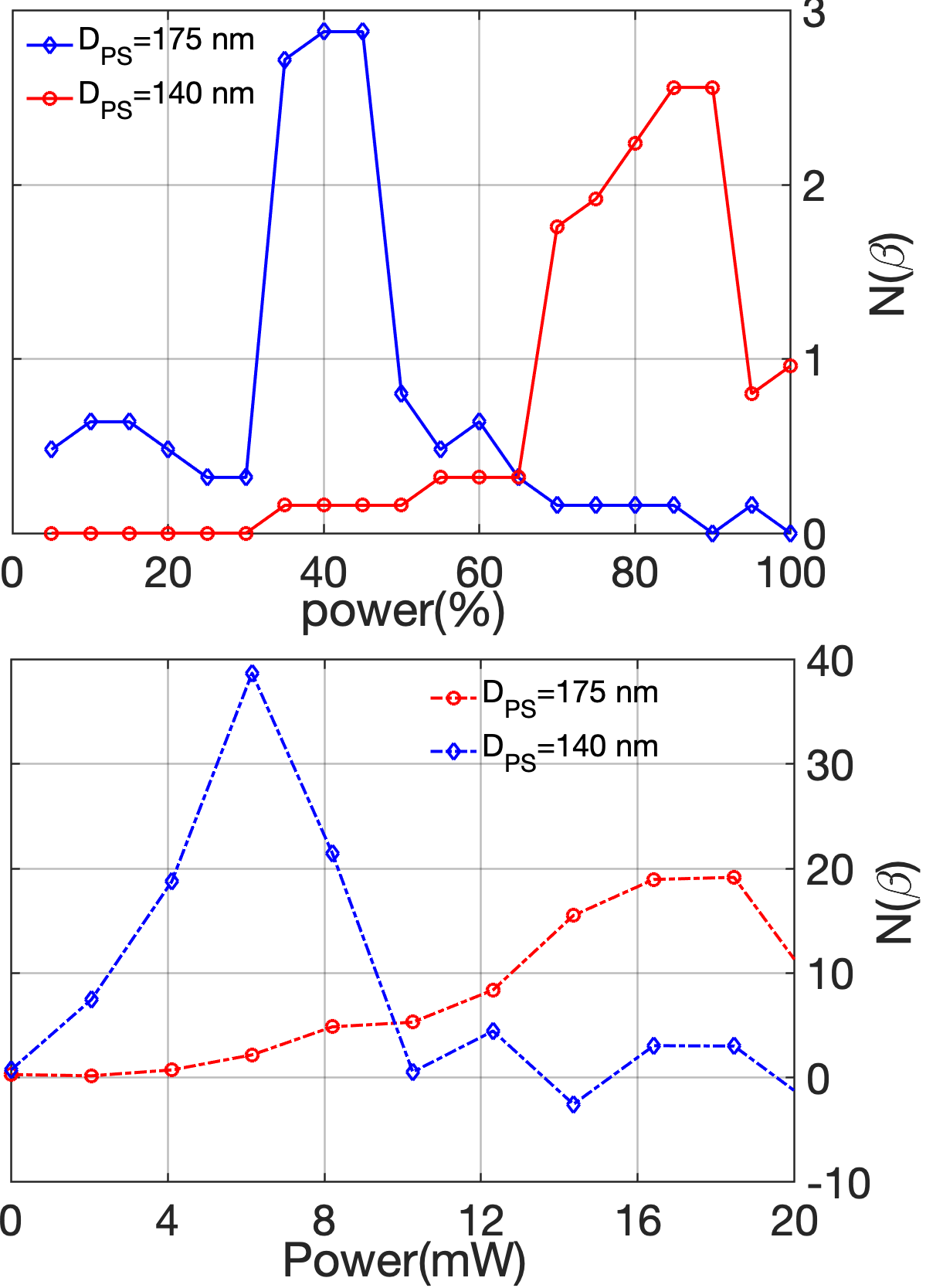}  
  \caption{}
  \label{fig:PS_dT}
\end{subfigure}

\begin{subfigure}{.3\textwidth}
  \centering
  \includegraphics[width=1\linewidth]{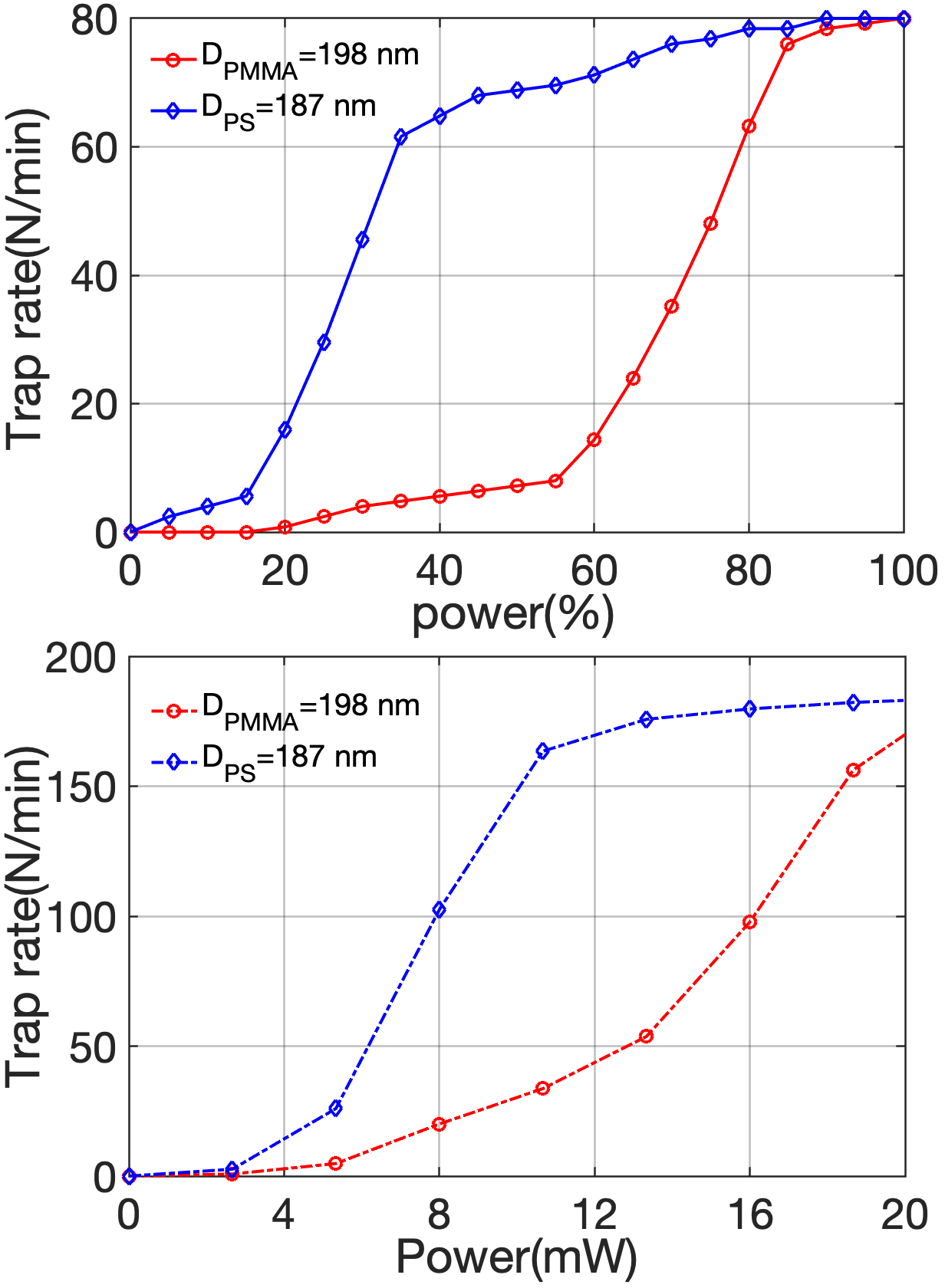}  
  \caption{}
  \label{fig:PM_T}
\end{subfigure}
\begin{subfigure}{.29\textwidth}
  \centering
  \includegraphics[width=1\linewidth]{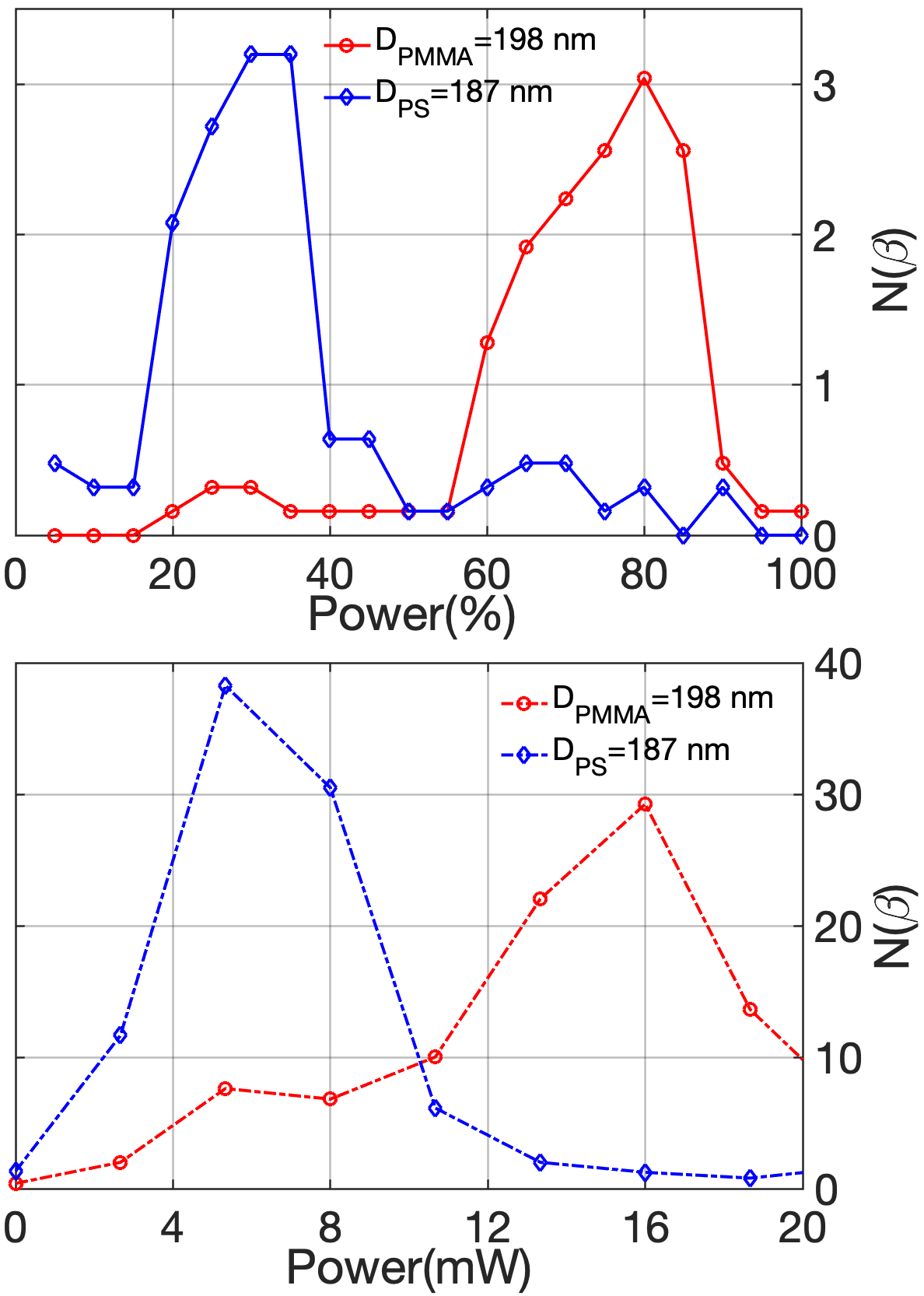}  
  \caption{}
  \label{fig:PM_dT}
\end{subfigure}

\caption{(a-d) Experimental results  (upper row) versus simulation results (lower row) when polystyrene (PS) nanoparticles with diameters of 140~nm, 175~nm, 187~nm and PMMA nanoparticles with diameter of 198~nm are introduced through the cavity.}
\label{fig:PS_PMMA}
\end{figure}

Figure \ref{fig:PS_PMMA}(a-d) compares experimental (upper row) and simulated (lower row) data for the trapping behaviour as a function of laser power, with each panel comparing two different nanoparticle samples. Sub-figures (a) and (c) show raw data in which the laser power is swept and the rate of trap events is recorded, while sub-figures (b) and (d) show the derivative of these raw data with respect to power which represents a measure of the distribution $N(\beta)$.

The experimental data in figures \ref{fig:PS_PMMA}a and \ref{fig:PS_PMMA}c show clear steps in trap rate with laser power, agreeing well with the simulations. The effect of the parabolic flow can be seen by a small increase in trap rate at laser powers below the main step, consistent with the lower threshold of trapping for the slower moving particles close to the mirrors.

Figure \ref{fig:PS_PMMA}a and \ref{fig:PS_PMMA}b show that PS particles of radius 140~nm and 175~nm are clearly resolved by the technique. The 140~nm particles require about double to power of the 175~nm particles to be trapped, such that the threshold power for trapping scales as $\sim a^3$ in contrast to equation \eqref{beta}. This dependency suggests that these experiments are in a regime where the trap time is determined more by the thermal energy than by the fluid flow. 
The full-width-at-half-maxima of the peaks in figure \ref{fig:PS_dT} reveals a sizing resolution of about 10~nm. Figure \ref{fig:PS_PMMA}c and \ref{fig:PS_PMMA}d show similarly that PS ($n_p$ = 1.59) particles of radius 187~nm and PMMA ($n_p$ = 1.49) particles of radius 198~nm are comfortably distinguished, the PMMA particles requiring a high laser power from trapping despite being slightly smaller. The full-width-at-half-maxima of the peaks in \ref{fig:PS_PMMA}b reveals a refractive index resolution of about 0.03 RIU.

We note that the resolution of the technique could in principle be increased significantly by increasing both the intracavity power and flow speed. An estimate of the quality factor of the measurement is provided by the optical trap strength: the ratio of the depth of the trap to the thermal energy,
\begin{equation}
    Q=\frac{\alpha P}{\pi n c w^2 K_B T}.
\end{equation}
For the experiments presented here, $Q \sim 3$, with $P$ limited to around 20 mW by intracavity heating effects. These occur at a relatively low average power due to the measurement method which sweeps the cavity mode through resonance with the laser, such that the peak power in the cavity is some fifty times greater than the effective $P$ for trapping. We now briefly consider other approaches that could be taken in which the trapping laser is `always on' such that the peak power and average power are equal. Equation \eqref{E:beta_thresh} reveals that a factor of 50 increase in $P$ would lead to a concomitant reduction in $\beta_T$, such that particles about seven times smaller would be trapped. Equivalently a larger quality factor $Q \sim 100$ could potentially be achieved by simultaneously increasing the flow speed to around 1.5 mm s$^{-1}$ to provide a resolution to particle size of below 1 nm or to refractive index of order 10$^{-3}$ RIU. An additional benefit of the increased flow speed would be that each trap measurement would take only a few milliseconds to record.

While based on single particle measurements, the technique as presented provides only ensemble data in the form of the distribution $N(\beta)$. Adaptations could in principle measure $\beta$ for each particle, potentially by sweeping the trap power during the trapping event. Such developments would be valuable in extending the range of parameters that can be measured on single particles. For example the maximum mode shift provides a measure of the polarisability of each particle \cite{trichet2016nanoparticle} and so could be combined with $\beta$ to yield both particle radius and refractive index.

The measured distribution $N(\beta)$ provided by this method can be used to establish sample-to-sample variations in particle size and composition, complementary with other analytic techniques. Since the data recorded contains further information beyond the duration of trap events it is also possible to use this method as part of a wider technique to provide independent measurements of size and composition parameters.

Additional improvements might be achieved by pumping the particles through the cavity using dielectrophoresis, which is capable of establishing a uniform flow rate across the flow channel. The technique might also be used to measure thermophoretic effects that result from particle heating and can provide information on particle thermal conductivity \cite{mcnab1973thermophoresis,schermer2011laser} or ionic shielding \cite{duhr2006molecules}.

\pagebreak
\begin{suppinfo}
\renewcommand{\theequation}{S.\arabic{equation}}
\renewcommand{\thefigure}{S.\arabic{figure}}
\setcounter{equation}{0}
\setcounter{figure}{0}
\subsection{Gaussian beam parameters}

The plano-concave optical microcavity is characterised by the cavity length $L$ and the radius of curvature of the concave mirror $R_L$. The Rayleigh range $z_R$ of the confined mode is then
\begin{equation}\label{zR}
\begin{aligned}
z_R=L\sqrt{\left(\frac{R_L}{L}-1\right)}.
\end{aligned}
\end{equation}
The radius of curvature of the wavefronts $R(z)$ and the beam radius $w(z)$ are then given by

\begin{equation}\label{E:Rz}
R(z)=z\left(1+\left(\frac{z_R}{z}\right)^2\right),
\end{equation}
\begin{equation}\label{E:Wz}
w(z)=\sqrt{\frac{2zR(z)}{kz_R}}.
\end{equation}
 
 The intensity distribution for the resonant cavity mode supported between opposing concave and planar mirrors given in equation \eqref{E:I_R} is determined as the sum of two Gaussian beams propagating in opposite directions, $I=n_{m}\epsilon_0c\lvert{E_i+E_r}\lvert^2$, where the incident ($E_i(\rho,z)$) and reflected ($E_r(\rho,z)$) waves are given by \cite{zemanek1998optical}:

\begin{equation}\label{E:E_i}
\begin{aligned}
E_i(\rho,z)=E_0\frac{\omega_0}{\omega}\exp(\frac{-\rho^2}{\omega^2})\exp(ikz+\frac{ik\rho^2}{2R}-i\arctan(\frac{z}{z_R})),
\end{aligned}
\end{equation}
\begin{equation}\label{E:E_r}
\begin{aligned}
E_r(\rho,z)=E_0\frac{\omega_0}{\omega}\exp(\frac{-\rho^2}{\omega^2})\exp(-ikz-\frac{ik\rho^2}{2R}+i\arctan(\frac{z}{z_R})).
\end{aligned}
\end{equation}

For the measurements reported here we used $R_L=12~\mu$m and $L=960$ nm.
\begin{figure}[!t]
\centering
    \includegraphics[width=.6\textwidth]{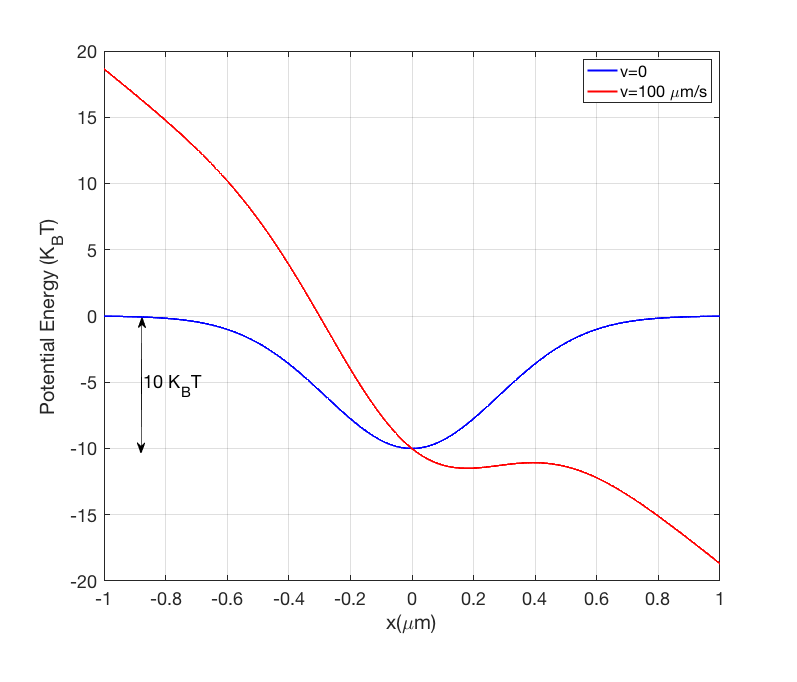}
\caption{Total potential energy along the axis $(x,0,0)$ for a PS particle radius of 100 nm with no flow (blue curve) and with a flow speed of 100 $\mu$m s$^{-1}$ (red curve).}
\label{fig:potential_energy}
\end{figure}

\subsection{Balancing optical and flow forces}
The maximum optical force is obtained by establishing the maximum value of $\frac{dI}{dx}$ using the mode intensity in equation \eqref{E:I_R}. This maximum force opposing the drag force due to the fluid flow occurs at $x=\frac{w}{2}, y=0, z=0$ at which point $\frac{dI}{dx}=\frac{2P}{\pi w^3 \sqrt{e}}$. Substitution of this expression into the optical force term in equation \eqref{Brownian} and equating with the viscous drag force due to fluid flow yields equation \eqref{E:beta_thresh}. This condition corresponds approximately to the potential energy profile shown in red in figure \ref{fig:potential_energy}, where the potential gradient tends to zero at the downstream side of the cavity mode. 

\subsection{Monte Carlo simulation}

Based on Equations \eqref{Brownian} and \eqref{E:I_R}, and the selection of the $x$ axis as the flow, the equations for incremental movements of a particle in the Cartesian coordinate system are:
\begin{equation}\label{browninanx}
\begin{aligned}	
x_{i}= x_{i-1}+\frac{\Delta t}{2n_m \gamma \epsilon_0 c}\alpha \frac{dI\left(\underline{r}\right)}{dx}+
\sqrt{\frac{2K_B T \Delta t}{\gamma}}~w_{i} 
+ \Delta t {v_x}_0,
\end{aligned}
\end{equation}
\begin{equation}\label{browninay}
\begin{aligned}	
y_{i}= y_{i-1}+\frac{\Delta t}{2n_m \gamma \epsilon_0 c}\alpha \frac{dI\left(\underline{r}\right)}{dy}+
\sqrt{\frac{2K_B T \Delta t}{\gamma}}~w_{i},
\end{aligned}
\end{equation}
\begin{equation}\label{browninaz}
\begin{aligned}	
z_{i}=z_{i-1}+\frac{\Delta t}{2n_m \gamma \epsilon_0 c}\alpha \frac{dI\left(\underline{r}\right)}{dz}+
\sqrt{\frac{2K_B T \Delta t}{\gamma}}~w_{i}.
\end{aligned}
\end{equation}
 Here $w_i$ is a computer generated, normally distributed random number with unity variance. The time increment $\Delta t$ is selected as 1 $\mu$s which is short enough to prevent `tunneling' of the particle through the potential barriers. 
The Monte Carlo model allows simulation of the mode shift with time as the particle moves through the mode $I\left(r\right)$. 
\begin{figure}[!t]
	\centering
	\begin{subfigure}[b]{.33\textwidth}
		\centering
		\includegraphics[width=1\linewidth]{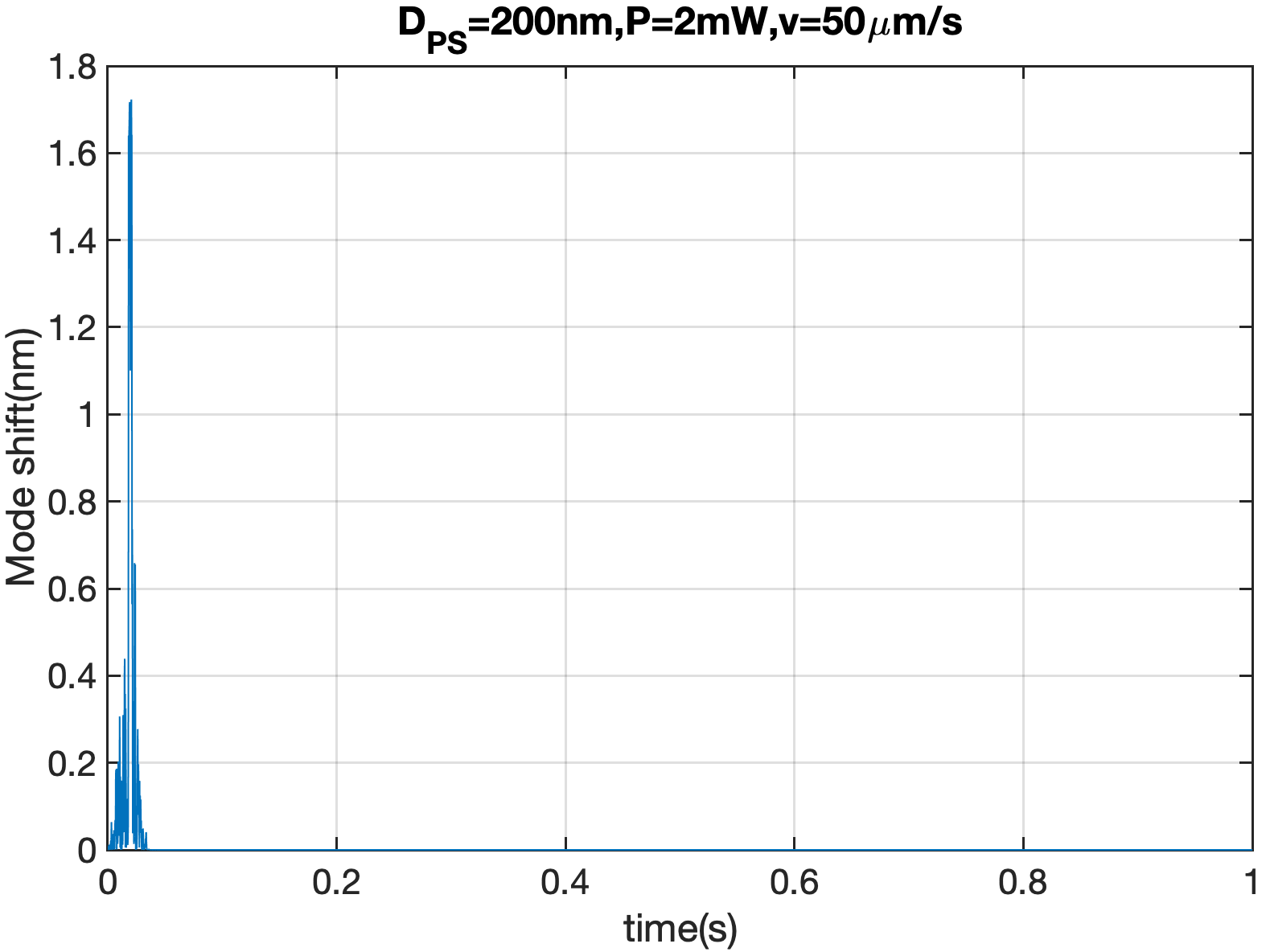}
		\caption{~}
		\label{fig:modeshift_diffused}
	\end{subfigure}
	\begin{subfigure}[b]{.33\textwidth}
		\centering
		\includegraphics[width=1\linewidth]{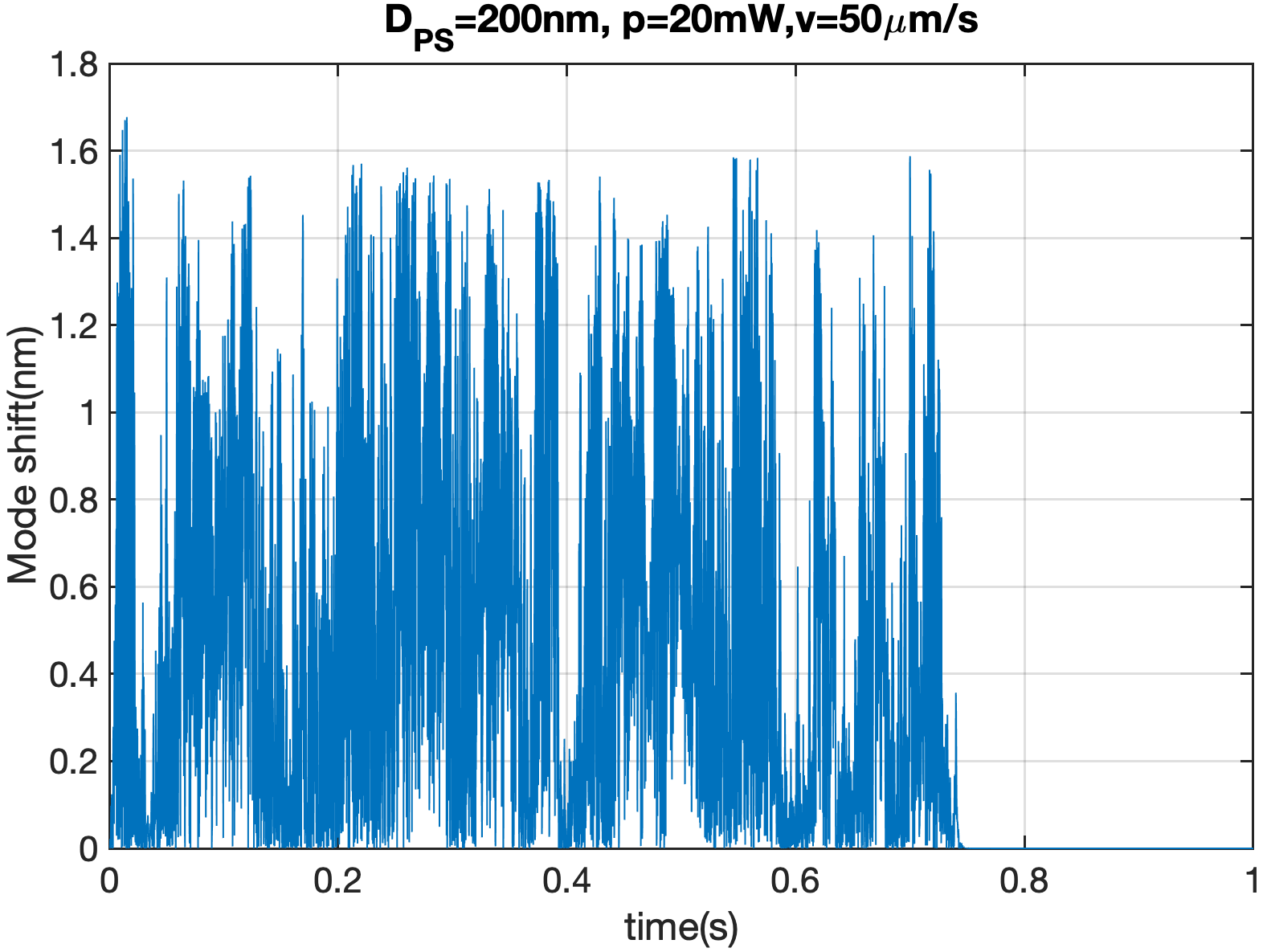}  
		\caption{~}
		\label{fig:modeshift_trapped}
	\end{subfigure}
	\caption{The mode shift of a PS particle with radius of 100 nm with a fluid flow speed of 50 $\mu$m s$^{-1}$ and intracavity powers of (a) 2 mW and (b) 20 mW.}
	\label{fig:modeshift_simulation}
\end{figure}
Figure \ref{fig:modeshift_diffused} shows example mode shift events of a spherical PS nanoparticle diffusing through the cavity. The radius and the velocity of the nanoparticle are 100 nm, and 50 $\mu$m s$^{-1}$, respectively. At an intracavity power of 2 mW the particle passes through the cavity mode in about 20 ms while at an intracavity power of  20 mW the particle remains in the mode for about 750 ms (figure \ref{fig:modeshift_trapped}).


\end{suppinfo}


\bibliography{References}

\end{document}